\documentstyle[12pt]{article}
\begin{document}
\begin{titlepage}
\begin{flushright}
{\bf HU-SEFT R 1994-06}
\end{flushright}
\begin{center}
{\Large\bf Conserved Quantities and Electroweak Phase Transitions}
\vskip 2.0 cm
{\bf M. Chaichian}\renewcommand{\thefootnote}{1}\footnote{Laboratory of High
Energy Physics , Department of Physics, P.O. Box 9
(Siltavuorenpenger 20C) FIN-00014, University of Helsinki, Finland}$^,$
\renewcommand{\thefootnote}{2}\footnote{
Research Institute for High Energy Physics, P.O. Box 9 (Siltavuorenpenger
20C) FIN-00014, University of Helsinki, Finland} ,
{\bf R. Gonzalez Felipe}$^{1,*}$
 and
{\bf H. Perez Rojas}$^{2,}$\renewcommand{\thefootnote}{3}\footnote{
Research Institute for Theoretical Physics, P.O. Box 9 (Siltavuorenpenger
20C) FIN-00014, University of Helsinki, Finland}$^,$
\renewcommand{\thefootnote}{*}\footnote{On
leave of absence from Grupo de F\'{\i}sica Te\'orica, ICIMAF,
Academia de Ciencias de Cuba, Calle E No. 309,Vedado, La Habana 4, Cuba}

\end{center}
\vskip 4.0 cm
\begin{abstract}

Some cosmological consequences of including the adequate conserved
quantities in the density matrix of the electroweak theory
are investigated. Several arguments against including the
charges associated to the spontaneously broken symmetry are presented.
Special attention is focused on the phenomenon of $W$-boson condensation
and its interplay with the phase transition for the symmetry restoration is
considered. The emerging cosmological implications, such as on the
baryon and lepton number densities, are of interest.
\end{abstract}
\end{titlepage}

After the  suggestion made by Kirzhnitz and Linde ~\cite{kirz} about the
mechanism of symmetry
restoration by temperature, electroweak phase transitions became a subject of
continuous interest (see, e.g. ~\cite{quir}). Among
other reasons, this interest has increased in the past few years
 after the observation that nonperturbative processes (mediated by
sphalerons) are relevant to the problem of the baryon asymmetry of the
Universe \cite{{kuzm},{shap}}.

In this letter we address some specific features which are essential in
the formulation  of the basic statistical theory underlying the thermodynamics
of the standard electroweak model,
in which the notions of gauge invariance, spontaneously broken
symmetries and
Higgs mechanism play a basic role. We shall consider the consequences
of including the adequate conserved charges, together
with the Hamiltonian, in the density matrix. When studying
the electroweak phase transitions the canonical ensemble is the one
usually
used. The latter, however, restricts the domain of physical applications,
especially as one would like to incorporate the baryon asymmetry
of the Universe and the equilibrium among electrically charged particles.
In some previous papers ~\cite{{kala},{ferr}}, it has been pointed out
 that the grand canonical
ensemble must be used. Moreover, as emphasized in ~\cite{chaic} ,
the conserved quantities to be included must be essentially
the electric charge and the lepton and baryon numbers. Thus, the
weak neutral charge associated with the spontaneously broken symmetry
 must be excluded.

Here we would like to discuss the above-mentioned points in connection with
a new phase transition which arises when these charges are
included, namely the $W$-vector boson condensation (which for high
temperatures occurs at very low lepton densities) ~\cite{{kala},{chaic}}
 and its interplay
with the phase transition of symmetry restoration, which according
to the recent results ~\cite{carr} is of first order if the perturbative
expansion is taken beyond the one-loop approximation.

The basic idea in \cite{kirz} came out from the analogy between the
Weinberg-Salam Lagrangian and the Ginzburg-Landau free energy for
 superconductivity
which is regarded as a condensate of Cooper pairs. In the electroweak case,
 the
role of the superconductor ground state is played by the vacuum whose symmetry
has been spontaneously broken. This breaking of the
symmetry is then understood in a hot Universe as a consequence of a phase
transition by cooling below some critical temperature $T_c$.

One could be tempted to interprete the asymmetric vacuum of electroweak
 theory
as a weak-neutrally charged condensate of scalar particles. However, this
picture
is not correct once it mixes the concepts of statistics (many-particle
systems)
with those of quantum field theory. By relativistic invariance we must
require that for the vacuum $P_\mu |0 \rangle = 0$, i.e.  the
four-momentum vector must have zero eigenvalue. Obviously, a condensate of
 scalar particles would violate this condition.

In a correct interpretation, however, the vacuum is asymmetric at any
temperature and
the symmetry breakdown is characterized by some parameter $\xi = a$, which is
the minimum of an effective potential in quantum field theory. By heating ,
 the effective
field amplitude (related to the nonzero vacuum charge, see below) is
decreased to the value  $\xi^2 (T) = a^2 - \alpha T^2$
due to screening, entropy-increasing processes, coming from
the temperature quantum loop corrections . At
the critical temperature $T_c$ we obtain $\xi^2 (T) = 0$ as the solution
corresponding to
a minimum of the effective temperature-dependent potential. However,
this minimum should not
be understood as a new vacuum, but as a symmetric ground
state $|\psi \rangle$ of a many-particle system.

Two problems closely related to the above are the questions about the infrared
stability of the system (if
long-range weak forces appear after the symmetry restoration)  and the
conservation of the weak neutral charge $Q^N$. (Both were considered in
{}~\cite{kirz} and the second one, mentioned in ~\cite{wein}). If the
neutrality of the system with respect to the weak neutral charge,
 i.e. $\langle Q^N \rangle = 0$, can be defined at all,  then it would
be reasonable to require it only in the
symmetric phase in order to prevent long-range repulsive forces from
making the system unstable.  This argument apparently
becomes weakened if some small infrared mass of order $g^2 T$ appears
\cite{poly}.
Thus it is usually accepted that $\langle Q^N \rangle$ is a
 conserved quantity  only
for $T > T_c$. One should understand this conservation in a statistical and
not quantum
mechanical sense, which means that $Tr(\rho \dot{Q}^N) = Tr(i \rho [H,Q^N])
= f(\xi(T),T)$ vanishes for $T > T_c$, without the requirement
$ [H,Q^N] = 0$.
However,  this conservation does not occur below $T_c$ and moreover
 $Q^N$ becomes in quantum field theory an ill-defined operator \cite{{gale},
{chai}}. Indeed, according to Coleman's theorem \cite{cole}
a charge operator $Q$ annihilates the vacuum, i.e.
$Q|0\rangle = 0$,
is conserved, $[H,Q] = 0$. However, a charge associated with a spontaneously
broken symmetry
like $Q^N$ does not annihilate the vacuum,
$Q^N|0\rangle \neq 0 $ \cite{chai}.
As a consequence, the operator $Q^N$ has many odd features: one cannot
construct states of definite charge number and even define the properties of
 conjugation and composition law . Furthermore,
it has been stated that such an operator is non-Hermitean
\cite{gale} and
for finite energy conditions, even non-conserved at the classical level
\cite{chai}. The above
arguments  forbid the inclusion of $Q^N$
into the density matrix via a chemical potential. This can be understood
also from the facts that such a charge does not lead to
irreducible representations of states with
{\it definite} charge number in specific Hilbert spaces  once it mixes
representations \cite{chai}, and also that it forbids
superselection rules.

A fundamental quantum-statistical statement is that equilibrium states
satisfy the  so-called Kubo-Martin-Schwinger (KMS) condition
\cite{{arak},{haag}},
which requires that the density matrix $\rho$ as well as
$H' = H - \sum_i \mu_i N_i$ have a discrete
spectrum. The KMS condition is  obtained from the time translations of
any operator $A$ given by
$e^{iH't} A e^{-iH't}$, with the transformation which
involves the chemical
potentials being a $U(1)$ gauge transformation. The introduction of $Q^N$
 would
lead to non-unitary implementable transformations invalidating the KMS
condition.

The above arguments,
applied to the grand canonical ensemble for the electroweak model,
imply that only those charges that belong to the center of
 the unbroken
part of the $SU(2)_L \times U(1)$ group satisfy the KMS condition and
thus can  have a chemical potential \cite{haag}. In other
words, only baryon and lepton numbers as well as electric charge can be
included in the density matrix via their corresponding chemical potentials.
It could be argued that in the case of electric charge, the chemical potential
added to the fourth component of the charged gauge field
can be gauged away. However, since there is only an {\it analogy} and not
an {\it equivalence} between quantum statistics and quantum field theory via
the Bloch equation and the Matsubara formalism, this chemical potential
is to be understood at most as the analog of a constant background field,
not subject to gauge transformations (see e.g. \cite{boul}) .

We should emphasize that if one does what one should not do, and includes
 a chemical potential $\mu_3$ corresponding to the spontaneously broken weak
neutral charge, then one arrives at drastic
changes in the consequences of the electroweak theory applications in
cosmology \cite{chaic}. In particular, if such a chemical potential is
introduced \cite{lind}, one obtains a relation between the critical temperature
for the symmetry restoration and the lepton density, which in turn relates the
latter to the photon density. This then leads to the prediction that the lepton
density is by several orders of magnitude larger than the baryon density, a
belief which has been accepted starting from the work by Linde \cite{lind}.
According to our arguments such a chemical potential should not be included
and then one does not necessarily obtain the large abundance of lepton over
baryon densities. A preference for the latter situation has been advertised
\cite{olive} from different considerations and revisited cosmological models.
In fact, as we shall see below, a lepton density of the same order of magnitude
as the baryon density will emerge from the results of the present letter.

Taking into account the previous arguments, let us turn now our attention to
the problem of $W$-boson condensation, first pointed out by Linde \cite{linde}
for zero temperature but at finite densities.
In \cite{chaic} we have discussed the general formalism to be used in
order to obtain the thermodynamic
quantities. The effective potential $V(\xi)$, essentially
off-shell,
is adequate for the discussion of the symmetry restoration phase transition,
whereas
for obtaining the charge densities $N_i = -\partial \Omega /\partial \mu_i$,
the
thermodynamic potential $\Omega = V(\xi(T))$ should be used.
In \cite{kala},\cite{chaic} the phase diagram for $W$-boson condensation
was obtained assuming that the symmetry restoration phase transition was
second order. As mentioned above, it has been argued that such
transition can become first order if high-order terms in the loop
expansion are included. Therefore, it would be of interest to investigate also
the consequences of including such terms in the problem of $W$-condensation.
Recently,   the effective potential of the
standard model was obtained beyond the one-loop when perturbative expansions
include
ring diagrams \cite{fodo}. The problem of the coefficient of the term
linear in the temperature in the one-loop approximation is cured
by temperature-dependent radiative corrections and this term essentially
decides the nature of  the phase transition to be a first order one.

The conditions for condensation are obtained from the
infrared poles of the (gauge-dependent) $W$-boson Green function

\begin{equation}
\lim_{k_4 = 0,  \vec{k} \rightarrow 0} Det D^{-1}_{\mu \nu}(k_4 -i\mu_W,
\vec{k}, M_W) = 0 \ ,
\end{equation}
where $\mu_W$ is the chemical potential associated with the electric charge,
$M_W$ is the $W$-boson mass,
 $D^{-1}_{\mu \nu} = D^{-1}_{0\mu \nu}
+\Pi_{\mu \nu}$ ; $D^{-1}_{0\mu \nu}$ is the free propagator and
$\Pi_{\mu \nu}$ is the polarization operator. Eq. (1)
leads to different conditions for the condensation of transverse and
longitudinal
modes, the latter acquiring some extra mass due to the Debye screening
 \cite{fodo}. We have then

\begin{equation}
\mu_W^2 = M_W^2 \ , \ \ \mu_W^2 = M_W^2 + g^2 T^2\left(\frac{5}{6} +
\frac{n_f}{3}\right) \ ,
\end{equation}
where $g = e/\sin \theta_W$ is the $SU(2)$ coupling constant and
$n_f$ is the number of fermion generations. For simplicity we shall
consider in what follows only one generation of
leptons and quarks, i.e. $n_f = 1$.
As both equations cannot be
satisfied simultaneously, we conclude
that the condensation of the transverse modes is to be expected, which
leads
to the solutions $\mu_W = \pm M_W$. Which one
of them will be realized depends on the sign of the background charge.

In the one-loop approximation considered in \cite{chaic}, we found equal
masses for
longitudinal and tranverse modes, which was interpreted as a consequence of the
rotational invariance. However, the Debye mass for the longitudinal modes
appears already in the QED plasma without destroying the rotational invariance.
Instead of transverse and longitudinal modes, it would be better to speak
of magnetic and electric modes respectively, the tensor structure
of the both preserving
the rotational invariance. The condensation of the electrically charged
transverse modes has also an interesting analogy with what happens in
the QED plasma. If we consider
a very hot photon gas in equilibrium with a neutral electron-positron
background, we encounter the property that electric fields are screened
by the Debye term, but that it can exist in
equilibrium in the presence of a constant magnetic field; this is determined by
the infrared behavior of the photon Green function. We can understand then the
external magnetic field as a sort of condensate of transverse photons.
However, in our present case it is the charge and not the field which
condensates.

In what follows we take the critical temperature for $W$-boson condensation
(of the order of 100 GeV)
to be above the confinement temperature (of the order of 100 MeV), a situation
which is guaranteed when
the lepton density is sufficiently high (for a lower bound estimate of the
latter one can use Eq. (7)).

The asymptotic equilibrium equations  for electric charge,
lepton and baryon number conservation read as follows:

$$
\frac{\partial \Omega}{\partial \mu_W} = -\frac{2}{3} N_u + \frac{1}{3}N_d
+ N_e + N_W = 0 \ ,
$$

\begin{equation}
\frac{\partial \Omega}{\partial \mu_2} =  N_e + N_{\nu} = \ell \ ,
\end{equation}

$$
\frac{\partial \Omega}{\partial \mu_4} =  N_u + N_d = b \ ,
$$
where $\mu_2$
and $\mu_4$ correspond to lepton and baryon number conservation respectively
and  are introduced into the density matrix as factors
multiplying the corresponding
conserved Noether charges. In (3), $N_i , i=u, d, e, W, \nu,$ denote the net
number density of the corresponding particles minus their antiparticles.

 Asymptotically we have for fermions
\begin{equation}
N_{fL} = N_{fR} = \frac{\mu_f T^2}{6} \ ,
\end{equation}
while for $W$-bosons
\begin{equation}
N_W = \left(1 - \frac{g}{\pi} \sqrt{\frac{7}{6}}\right)\mu_W T^2 \ .
\end{equation}
{}From the interaction vertices we also have in the equilibrium the following
relations:
\begin{equation}
\mu_e = \mu_W + \mu_2 \; , \;
\mu_u = -2\mu_W/3 + \mu_4 \; , \; \mu_d = \mu_W/3 + \mu_4 \ .
\end{equation}

By solving Eqs. (3) and taking the condition of condensation for
the transverse modes as $\mu_W = - M_W = - g\xi(T)/2$
(i.e. taking the $\mu_W = - M_W$ solution in order to obtain the
situation $\ell - b/4 > 0$), we obtain finally
the critical curve for the condensation of transverse modes

\begin{equation}
 \ell -  \frac{b}{4} = - \frac{23}{12} \mu_W T^2  =
\frac{23}{24} g T^2  \xi(T) \ ,
\end{equation}
where the expression for $\xi(T)$ is obtained from the minimum of the
effective potential.  If the phase transition is of
second order, $\xi(T)$ decreases smoothly with increasing temperature,
and the phase
diagram has qualitatively the same form as the one obtained in \cite{chaic}.
 The more interesting case results when the nature of the phase transition is
 first  order .
Then in the process of cooling $\xi(T)$ jumps abruptly from zero to a
positive value with the formation
 of bubbles inside which $\xi(T) \neq 0$. The phase diagram
is depicted in Fig. 1.

We must point out here that the condensation of $W$-bosons partially diminishes
the
$m^3 T$ term needed for the first order phase transition to take place.
This is due to the fact that the $W$-bosons contribute to the effective
potential with a term of the form $(M_W^2 - \mu_W^2)^{3/2}T/(6\pi)$, which
vanishes if
the condensation arises. We believe, however, that the contributions of the
longitudinal and Higgs modes are enough to keep the linear in $T$ term
necessary for a first order phase transition.

Now if we assume that the effective density $\ell -  b/4$ in (7) is
large enough to preserve the condensate after the symmetry breaking, it is
easy to understand that the relative abundance of isotopic components of
leptons and quarks are strongly changed by the phase transition. Indeed,
we have the chemical equilibrium equations

\begin{equation}
\mu_u - \mu_d = \mu_\nu - \mu_e = - \mu_W = M_W \ .
\end{equation}
According to Eqs. (8), in the symmetric phase ($M_W = 0$)
there exists equal relative
abundance of $u$ and $d$ quarks, as well as of electrons and neutrinos. After
the breaking of the symmetry, the $W$-boson condensation which is induced by
the lepton density leads to a displacement of the equilibrium with a resulting
average excess of $u$ over $d$ quarks. This could  favour the
formation of an excess of
protons over neutrons with the cooling of the Universe below the confinement
critical temperature
(which corresponds to the present state of the Universe), if the processes
which
dominate at lower temperatures, such as nucleosynthesis, do not totally
erase this
asymmetry. We notice that identical to (8) equations are separately valid for
other lepton and quark generations and thus similar conclusions can be drawn
for them as well.

There might also exist a connection between the $W$-boson condensation
mechanism and a partial understanding of the baryon asymmetry of the Universe,
first treated by Sakharov in his pioneer work \cite{sakha} using a microscopic
as well as a thermodynamic approach. In the broken phase, according to Eqs.
(3) and (4), one would have $\mu_u + \mu_d = 3b/T^2$. With the occurrence of
$W$-condensation, according to Eq. (8) one could expect a value of order
$M_W T^2/3$ for $b$, leading to an enhancement of the baryon number density.
We would like to mention that similarly, from Eqs. (3) and (4), it follows
that the lepton number density $\ell$ is also expected to have a value of the
same order of magnitude as the baryon number density $b$, in contrast to the
result $\ell \gg b$ obtained in \cite{lind} in an equivalent way if the
spontaneously broken weak neutral charge would have been taken into account as
a conserved quantity. The equality, in the order of magnitude, of $\ell$ and
$b$ has been argued for \cite{olive} from entirely different cosmological
considerations.
We thus consider that
the phenomenon of  $W$-boson
condensation may have interesting cosmological implications.
\vskip 1.5 cm
\noindent
{\bf Acknowledgements}
\vskip 0.5 cm
We are grateful to J. Ignatius,
K. Kajantie, C. Montonen, V. Semikoz and M. Shaposhnikov for useful
discussions and comments.

\vskip 2.0 cm
\noindent
{\large\bf Figure captions}
\vskip 0.5 cm
\noindent
{\bf Fig. 1} \  Phase diagram for the high temperature $W$-condensation
 in the case of first-order phase transition.

\end{document}